\documentclass{appolb}
\usepackage{graphicx}

  \usepackage{amssymb}
   \usepackage{amsmath}
    \usepackage{amsfonts}
     \usepackage{epsfig}
      \usepackage{bm}
\usepackage{multirow}
\usepackage{ctable}
\usepackage{booktabs}
\usepackage{array}
\usepackage{tabularx}
\usepackage{xcolor}
\usepackage{pstricks}

\begin{document}
\title{Simultaneous production of D and B mesons%
\thanks{Presented at XXIV Cracow EPIPHANY Conference on Advances in Heavy Flavour Physics, 9-12 January 2018}%
}
\author{Rafa{\l} Maciu{\l}a
\address{Institute of Nuclear
Physics, Polish Academy of Sciences, Radzikowskiego 152, PL-31-342 Krak{\'o}w, Poland}
}
\maketitle
\begin{abstract}
We present results of our studies of double-parton scattering (DPS) effects in
simultaneous production of heavy flavour mesons (charm and bottom). We discuss production of charm-bottom and bottom-bottom
meson-meson pairs in proton-proton collisions at the LHC. The calculation of DPS mechanism is performed within factorized
ansatz where each parton scattering is calculated within $k_T$-factorization approach. The hadronization is done with the help of fragmentation
functions. For completeness we compare results for double- and single-parton scattering (SPS). The SPS components are also calculated in the $k_{T}$-factorization
with the help of KaTie Monte Carlo generator. As in the case of double charm production also here
the DPS dominates over the SPS, especially for small transverse momenta.
We present several distributions and integrated cross sections 
with realistic cuts for simultaneous production of $D^0 B^+$ and $B^+
B^+$, suggesting future experimental studies at the LHC.
\end{abstract}
\PACS{13.87.Ce,14.65.Dw}
  
\section{Introduction}

In the ongoing era of LHC a precise theoretical description of high-energy proton-proton collisions requires inclusion of multiple-parton interactions (MPI).
Over last years several experimental and theoretical studies of soft and hard MPI effects for different processes have been performed (see \textit{e.g.} Refs.~\cite{Astalos:2015ivw,Proceedings:2016tff}). A major part of those analyses concentrate in particular on phenomena of double-parton scattering (DPS). Usually, examination of DPS mechanism essentially needs dedicated experimental analyses and is strongly limited because of large background coming from standard single-parton scattering (SPS).

We have clearly shown in our previous papers that double open charm meson production $pp \to D D \!\; X$ is one of the best reaction to study hard double-parton scattering effects at the LHC \cite{Luszczak:2011zp,Maciula:2013kd,vanHameren:2014ava,vanHameren:2015wva}. In this case the standard SPS contribution is much smaller than the DPS one and the cross sections for the DPS at the LHC are predicted to be quite large, in particular, much larger than for other reactions previously studied in this context in the literature. Following these observations we have also carefully examined DPS effects in $pp \to \mathrm{4jets} \!\; X$ \cite{Maciula:2015vza,Kutak:2016ukc}, $pp \to D^{0} + \mathrm{2jets} \!\; X$ and $pp \to D^{0}\overline{D^{0}} + \mathrm{2jets} \!\; X$ \cite{Maciula:2017egq} reactions.  

Here, we wish to present results of similar phenomenological studies of
DPS effects in the case of associated open charm and bottom 
$pp \to D^{0} B^{^{+}} \!\; X$ as well as double open bottom $pp \to B^{+} B^{+} \!\; X$ production.
We extend results known from the literature both for $b \bar b b \bar b$ \cite{DelFabbro:2002pw} and 
$c \bar c b \bar b$ \cite{Cazaroto:2013fua} by including hadronization effects, the LHCb detector acceptance, and especially, by including single-parton scattering mechanism for a first time in a fully consistent way.

\section{A sketch of our approach}

\subsection{Single-parton scattering}

\begin{figure}[!h]
\centering
\begin{minipage}{0.32\textwidth}
 \centerline{\includegraphics[width=1.0\textwidth]{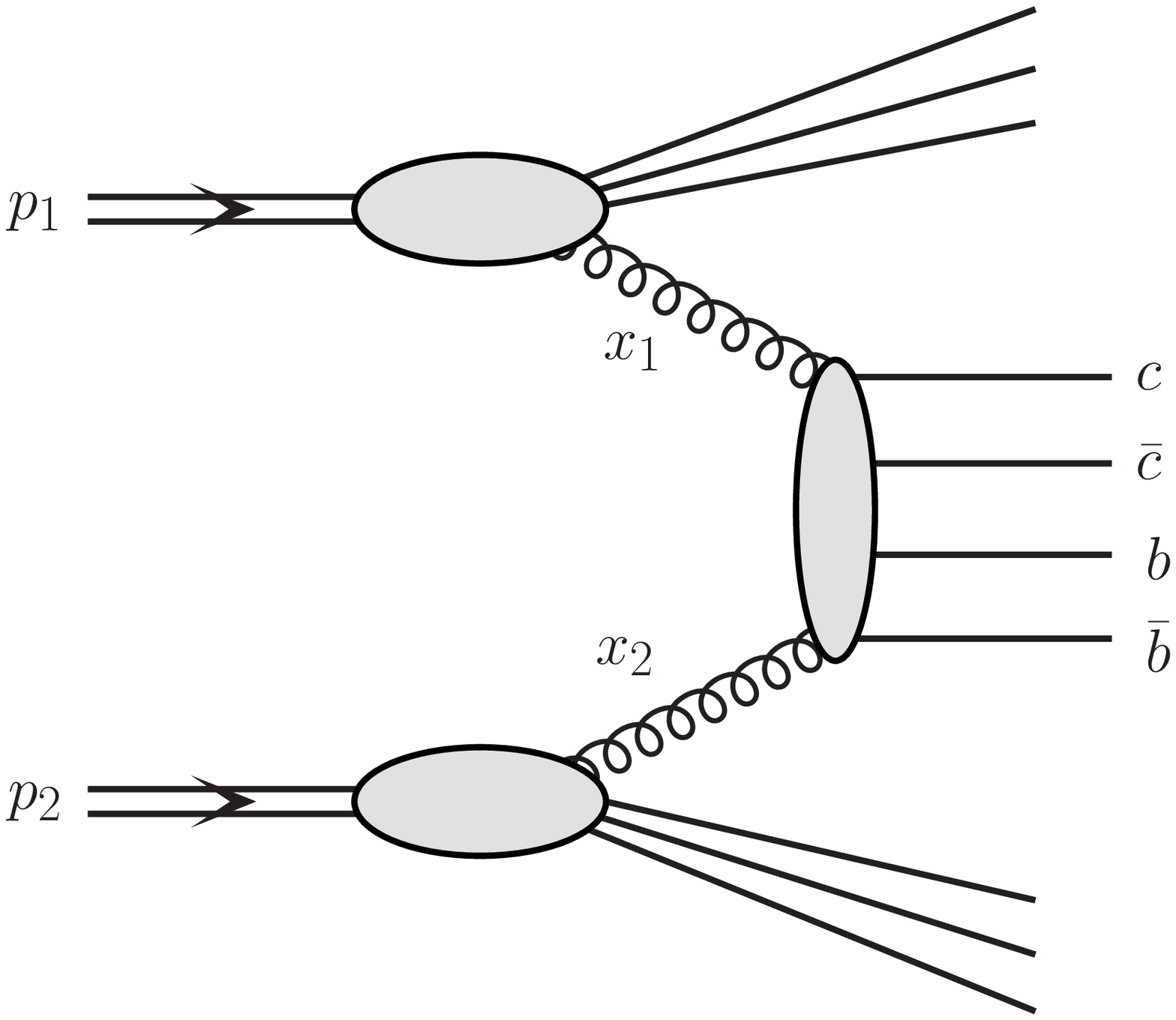}}
\end{minipage}
\hspace{0.1cm}
\begin{minipage}{0.32\textwidth}
 \centerline{\includegraphics[width=1.0\textwidth]{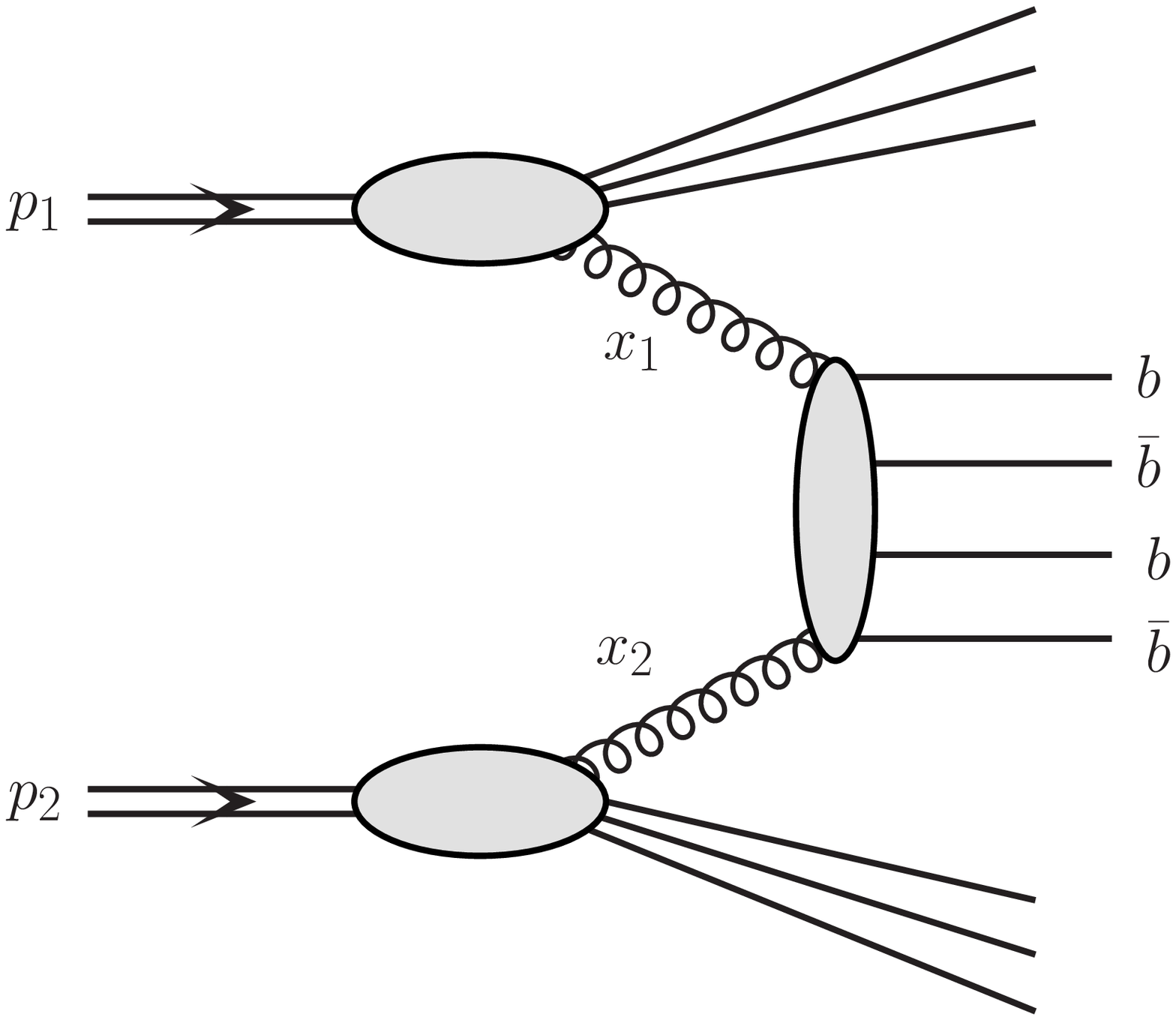}}
\end{minipage}
\caption{A diagrammatic representation of the dominant SPS mechanism
            for the $pp \to c\bar c b\bar b \, X$ (left panel) and for the $pp \to b\bar b b\bar b \, X$ (right panel) reactions.
\small 
 }
 \label{fig:diagrams_SPS}
\end{figure}

In the $k_T$-factorization approach \cite{Catani:1990xk,Catani:1990eg} the SPS cross section for $pp \to Q\bar Q Q\bar Q \, X$ reaction (see Fig.~\ref{fig:diagrams_SPS}) can be written as
\begin{equation}
d \sigma_{p p \to Q\bar Q Q\bar Q \; X} =
\int d x_1 \frac{d^2 k_{1t}}{\pi} d x_2 \frac{d^2 k_{2t}}{\pi}
{\cal F}_{g}(x_1,k_{1t}^2,\mu^2) {\cal F}_{g}(x_2,k_{2t}^2,\mu^2)
d {\hat \sigma}_{gg \to Q\bar Q Q\bar Q}
\; .
\label{cs_formula}
\end{equation}
In the formula above ${\cal F}_{g}(x,k_t^2,\mu^2)$ is the unintegrated
gluon distribution function (uGDF). The uGDF depends on longitudinal momentum fraction $x$, transverse momentum squared $k_t^2$ of the gluons initiating the hard process,
and also on a factorization scale $\mu^2$.
The elementary cross section in Eq.~(\ref{cs_formula}) can be written as:
\begin{equation}
d {\hat \sigma}_{gg \to Q \bar Q Q \bar Q } =
\prod_{l=1}^{4}
\frac{d^3 p_l}{(2 \pi)^3 2 E_l} 
(2 \pi)^4 \delta^{4}(\sum_{l=1}^{4} p_l - k_1 - k_2) \times\frac{1}{\mathrm{flux}} \overline{|{\cal M}_{g^* g^* \to Q \bar Q Q \bar Q}(k_{1},k_{2})|^2}
\; ,
\label{elementary_cs}
\end{equation}
where $E_{l}$ and $p_{l}$ are energies and momenta of final state heavy quarks.
The matrix element takes into account that both gluons entering the hard
process are off-shell with virtualities $k_1^2 = -k_{1t}^2$ and $k_2^2 = -k_{2t}^2$.
We limit the numerical calculations to the dominant at high energies gluon-gluon fusion channel for the $2 \to 4$ mechanism under consideration.

The off-shell matrix elements for higher final state parton
multiplicities are available, \textit{e.g.} through numerical methods based on BCFW recursion \cite{Bury:2015dla}.
Here, we use the KaTie code \cite{vanHameren:2016kkz}, which is 
a complete Monte Carlo parton-level event generator for hadron
scattering processes, for any initial partonic
state with on-shell or off-shell partons. 

In the numerical calculation, we use $\mu^2 \! = \! \sum_{i=1}^{4} m_{it}^{2}/4$ as the renormalization/factorization scale, where $m_{it}$'s are the transverse masses of the outgoing heavy quarks. We take running $\alpha_{s}$ at next-to-leading order (NLO),
charm quark mass $m_c$ = 1.5 GeV and bottom quark mass $m_b$ = 4.75 GeV. We use the Kimber-Martin-Ryskin (KMR) \cite{Kimber:2001sc,Watt:2003vf} unintegrated distributions for gluon calculated from the MMHT2014nlo PDFs \cite{Harland-Lang:2014zoa}. The effects of the $c \to D^{0}$ and $b \to B^{+}$ hadronization are taken into account via the scale-independent Peterson model of fragmentation function (FF) \cite{Peterson:1982ak}
with $\varepsilon_{c} = 0.05$ and $\varepsilon_{b} = 0.004$.  

\subsection{Double-parton scattering}

A textbook form of multiple-parton scattering theory (see \textit{e.g.}
Refs.~\cite{Diehl:2011tt,Diehl:2011yj}) is rather well established, however, not yet applicable for phenomenological studies. Generally, the DPS cross sections can be calculated with the help of the double parton distribution functions (dPDFs).
Unfortunately, the currently available models of the dPDFs are still rather 
at a preliminary stage. So far they are formulated only for gluon or for valence quarks and only in a leading-order framework.

Instead of the general form, one usually follows the so-called factorized ansatz, where the dPDFs are taken in the factorized form:
\begin{equation}
D_{1, 2}(x_1,x_2,\mu) = f_1(x_1,\mu)\, f_2(x_2,\mu) \, \theta(1-x_1-x_2) \, .
\end{equation}
Here, $D_{1, 2}(x_1,x_2,\mu)$ is the dPDF and
$f_i(x_i,\mu)$ are the standard single PDFs for the two generic partons in the same proton. The factor $\theta(1-x_1-x_2)$ ensures that
the sum of the two parton momenta does not exceed 1. 
%
\begin{figure}[!h]
\centering
\begin{minipage}{0.32\textwidth}
 \centerline{\includegraphics[width=1.0\textwidth]{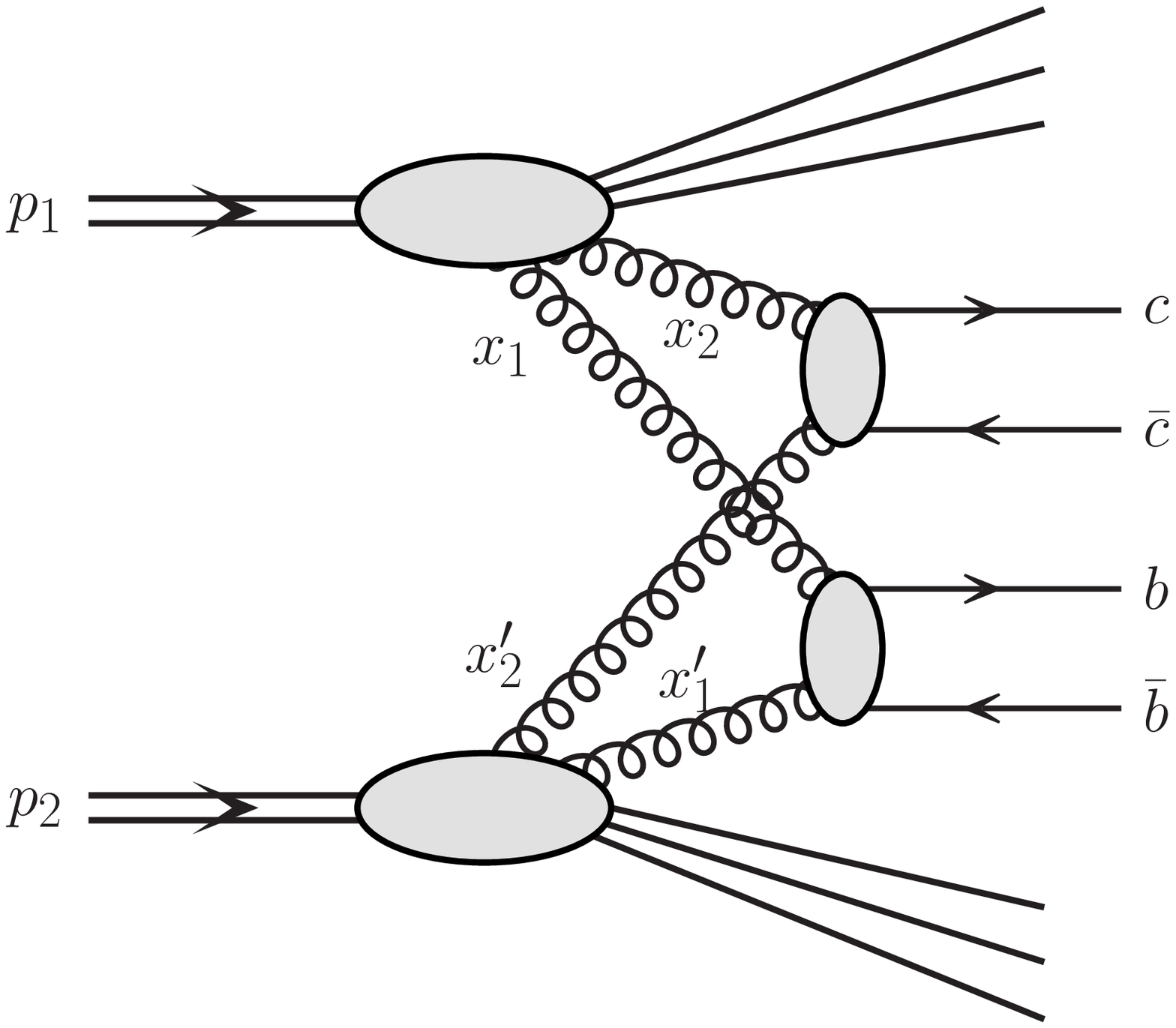}}
\end{minipage}
\hspace{0.1cm}
\begin{minipage}{0.32\textwidth}
 \centerline{\includegraphics[width=1.0\textwidth]{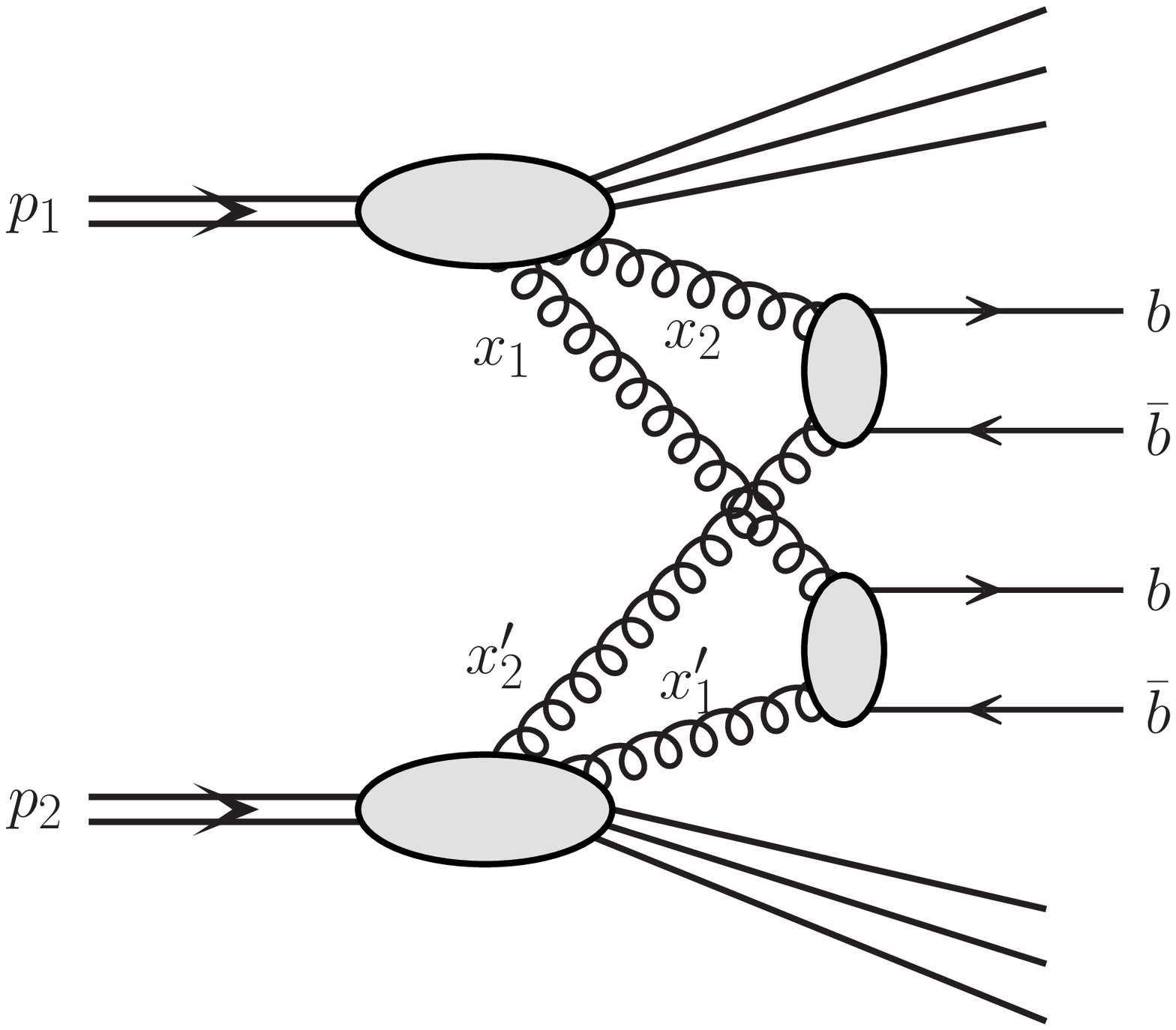}}
\end{minipage}
   \caption{A diagrammatic representation of the DPS mechanism
            for the $pp \to c\bar c b\bar b \, X$ (left panel) and for the $pp \to b\bar b b\bar b \, X$ (right panel) reactions.
\small 
 }
 \label{fig:diagrams_DPS}
\end{figure}

The differential cross section for $pp \to Q \bar Q Q \bar Q \; X$ reaction within the DPS mechanism, sketched in Fig.~\ref{fig:diagrams_DPS}, can be then expressed as follows: 
\begin{equation}
\frac{d\sigma^{DPS}(Q \bar Q Q \bar Q)}{d\xi_{1}d\xi_{2}} =  \frac{m}{\sigma_{\mathrm{eff}}} \cdot \frac{d\sigma^{SPS}(g g \to Q \bar Q)}{d\xi_{1}} \! \cdot \! \frac{\sigma^{SPS}(g g \to Q \bar Q)}{d\xi_{2}},
\label{basic_formula1}
\end{equation}
where $\xi_{1}$ and $\xi_{2}$ stand for generic phase space kinematical variables for the first and second scattering, respectively.
The combinatorial factor $m$ is equal $1$ for $c\bar c b\bar b$ and $0.5$ for $b \bar b b\bar b$ case. 
When integrating over kinematical variables one recovers the commonly used pocket-formula:
\begin{equation}
\sigma^{DPS}(Q \bar Q Q \bar Q) =  
m \! \cdot \!  \frac{\sigma^{SPS}(g g \to Q \bar Q) \! \cdot \! \sigma^{SPS}(g g \to Q \bar Q)}{\sigma_{\mathrm{eff}}}\; .
\label{basic_formula2}
\end{equation}

The effective cross section $\sigma_{\mathrm{eff}}$ provides normalization of 
the DPS cross section and is usually interpreted 
as a measure of the transverse correlation of the two partons inside 
the hadrons. The longitudinal parton-parton correlations are expected to be far less
important when the energy of the collision is increased. For small-$x$ partons and for low 
and intermediate scales the possible longitudinal correlations can be safely
neglected (see \textit{e.g.} Ref.~\cite{Gaunt:2009re}). In this paper we use world-average value of $\sigma_{\mathrm{eff}} = 15$ mb provided by 
several experiments at different energies (see \textit{e.g.} Ref.~\cite{Proceedings:2016tff} for a review).

In our present analysis cross sections for each step of the DPS
mechanism are calculated in the $k_T$-factorization approach, that is:
\begin{eqnarray}
\frac{d \sigma^{SPS}(p p \to Q \bar Q \; X_1)}{d y_1 d y_2 d^2 p_{1,t} d^2 p_{2,t}} 
&& = \frac{1}{16 \pi^2 {\hat s}^2} \int \frac{d^2 k_{1t}}{\pi} \frac{d^2 k_{2t}}{\pi} \overline{|{\cal M}_{g^{*} g^{*} \rightarrow Q \bar{Q}}|^2} \nonumber \\
&& \times \;\; \delta^2 \left( \vec{k}_{1t} + \vec{k}_{2t} - \vec{p}_{1t} - \vec{p}_{2t}
\right)
{\cal F}_{g}(x_1,k_{1t}^2,\mu^2) {\cal F}_{g}(x_2,k_{2t}^2,\mu^2),
\nonumber
\end{eqnarray}
\begin{eqnarray}
\frac{d \sigma^{SPS}(p p \to Q \bar Q \; X_2)}{d y_3 d y_4 d^2 p_{3,t} d^2 p_{4,t}} 
&& = \frac{1}{16 \pi^2 {\hat s}^2} \int \frac{d^2 k_{3t}}{\pi} \frac{d^2 k_{4t}}{\pi} \overline{|{\cal M}_{g^{*} g^{*} \rightarrow Q \bar Q}|^2} \nonumber \\
&&\times \;\; \delta^2 \left( \vec{k}_{3t} + \vec{k}_{4t} - \vec{p}_{3t} - \vec{p}_{4t}
\right)
{\cal F}_{i}(x_3,k_{3t}^2,\mu^2) {\cal F}_{j}(x_4,k_{4t}^2,\mu^2). \nonumber \\
\end{eqnarray}
The numerical calculations of the DPS are also done within the KaTie code.  
Here, the strong coupling constant $\alpha_S$ and uGDFs are taken the
same as in the case of the calculation of the SPS mechanism. The
factorization and renormalization scales for the two single scatterings
are $\mu^2 \! = \! \frac{m_{1t}^{2}+m_{2t}^{2}}{2}$ for the first, and 
$\mu^2 \! = \! \frac{m_{3t}^{2}+m_{4t}^{2}}{2}$ for the second subprocess.

\section{Numerical results}

\begin{figure}[!h]
\begin{minipage}{0.47\textwidth}
\centering
 \centerline{\includegraphics[width=1.0\textwidth]{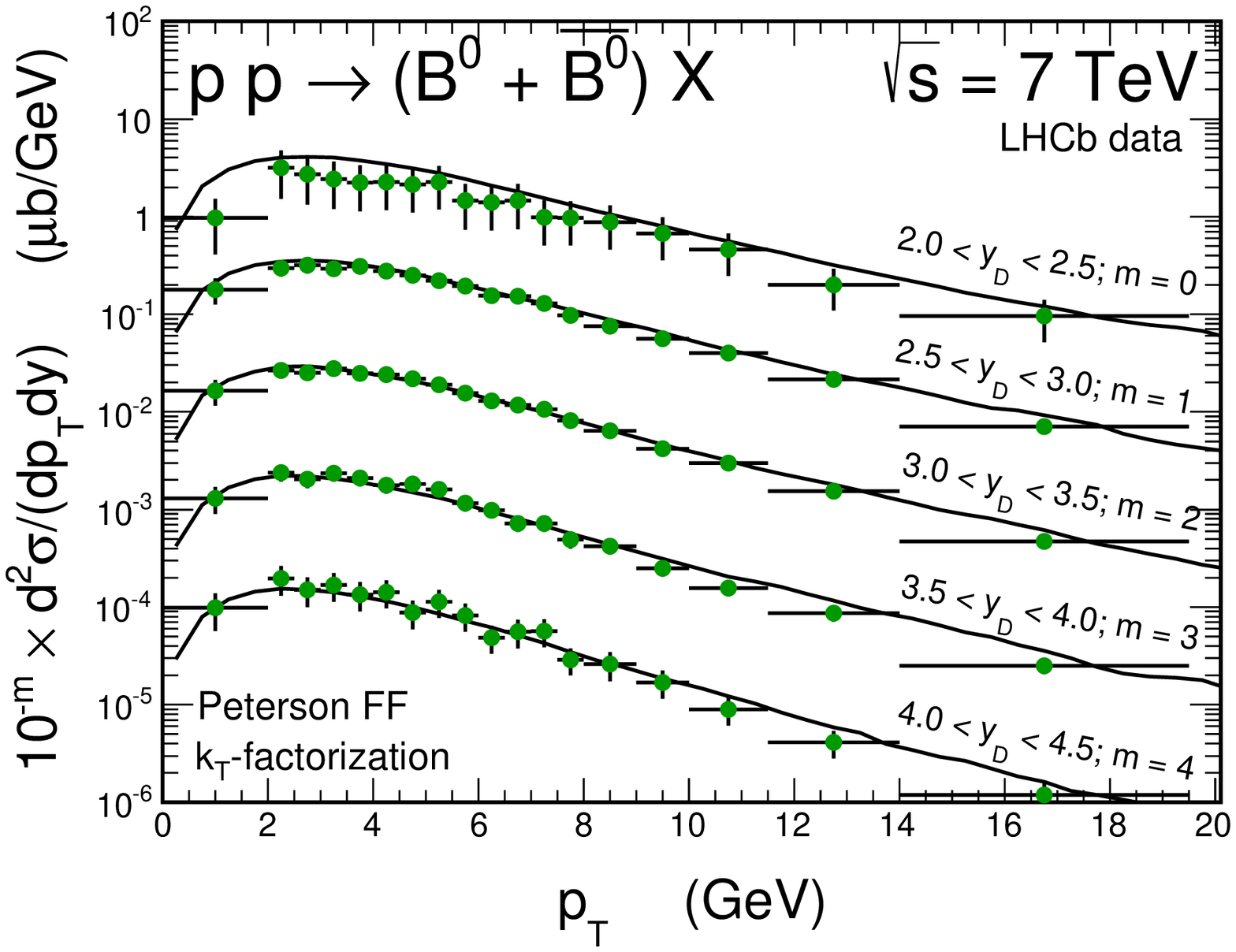}}
\end{minipage}
\hspace{0.5cm}
\begin{minipage}{0.47\textwidth}
 \centerline{\includegraphics[width=1.0\textwidth]{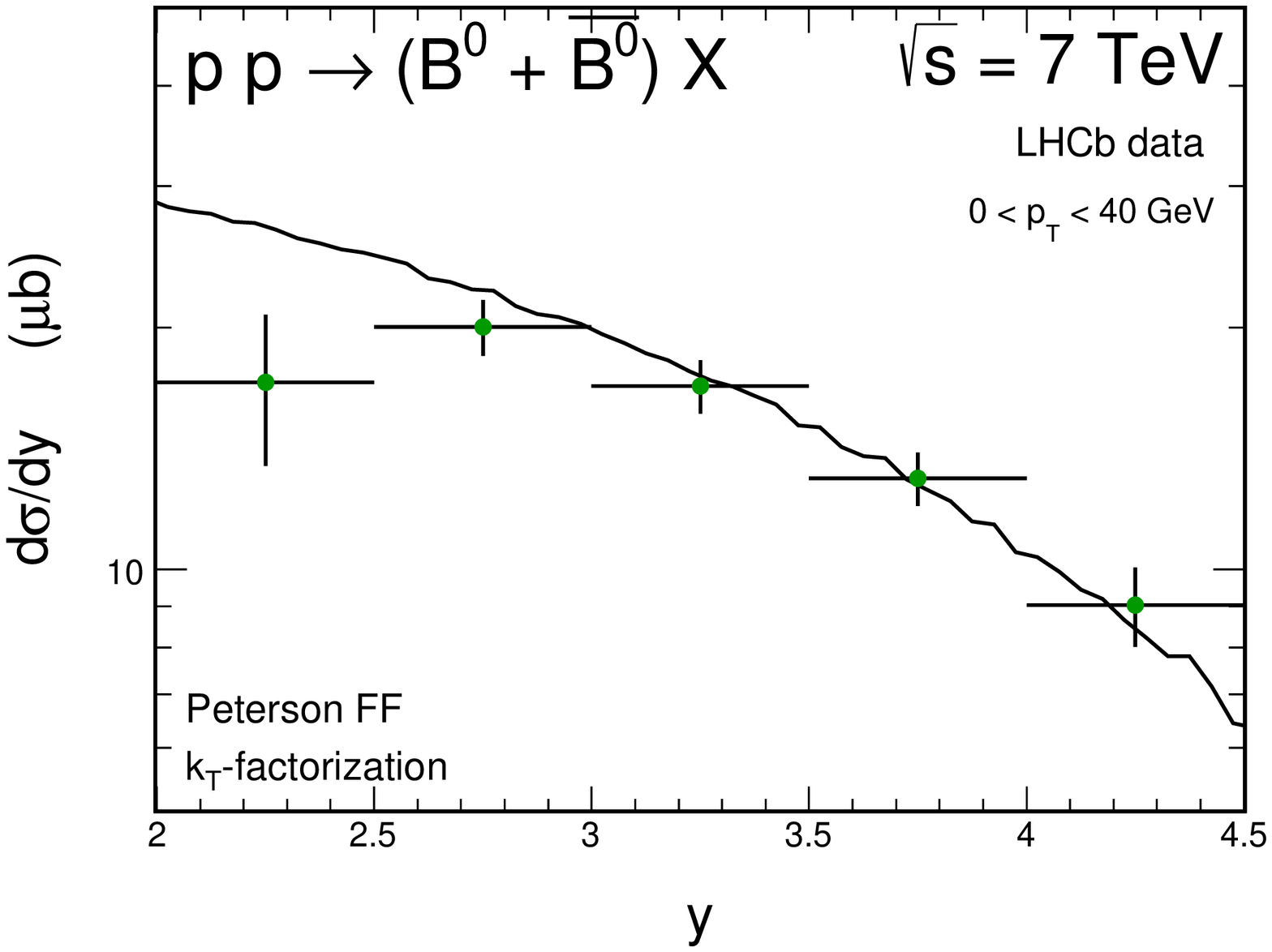}}
\end{minipage}
   \caption{
\small Transverse momentum (left) and rapidity (right) distributions of $B^{0}$ meson measured by the LHCb experiment at $\sqrt{s} = 7$ TeV \cite{Aaij:2013noa}. Theoretical predictions (solid lines) are calculated within the $k_{T}$-factorization approach with the KMR uPDFs. Details are specified in the figure.
 }
 \label{fig:B0}
\end{figure}

We start this section with presentation of results for inclusive open bottom meson production. In
Fig.~\ref{fig:B0} we compare our theoretical predictions based on the
$k_{T}$-factorization approach with the LHCb experimental data
\cite{Aaij:2013noa} at $\sqrt{s}=7$ TeV. We get a very good agreement
with the experimental points for both, the transverse momentum (left
panel) and rapidity (right panel) $B^{0}$ meson distributions. Only the
cross section in the lowest rapidity bin $y \in (2.0,2.5)$ seems to be
slightly overestimated, however the experimental uncertainties in this
case are noticeably larger than in other rapidity intervals.

\begin{figure}[!h]
\begin{minipage}{0.47\textwidth}
\centering
 \centerline{\includegraphics[width=1.0\textwidth]{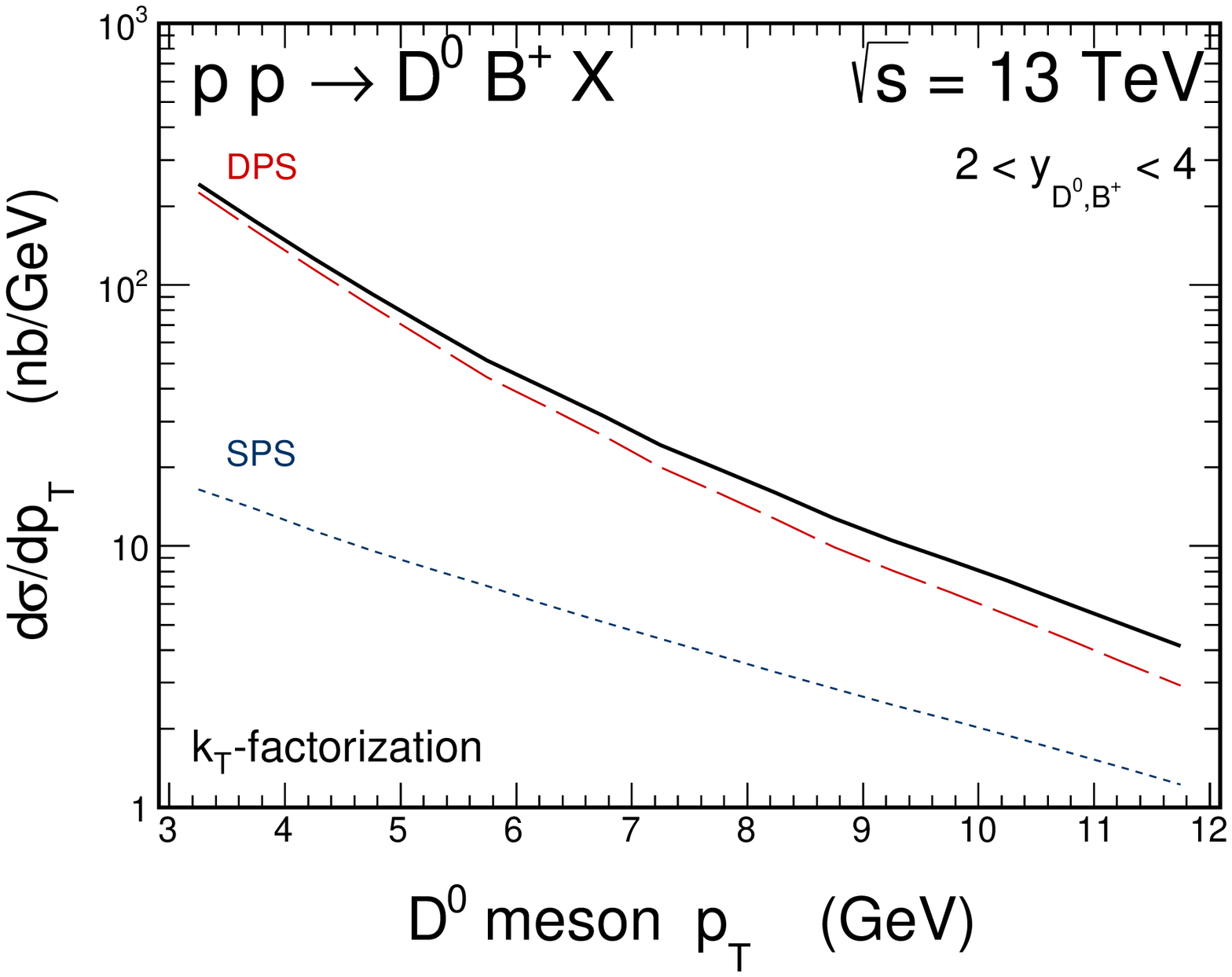}}
\end{minipage}
\hspace{0.5cm}
\begin{minipage}{0.47\textwidth}
 \centerline{\includegraphics[width=1.0\textwidth]{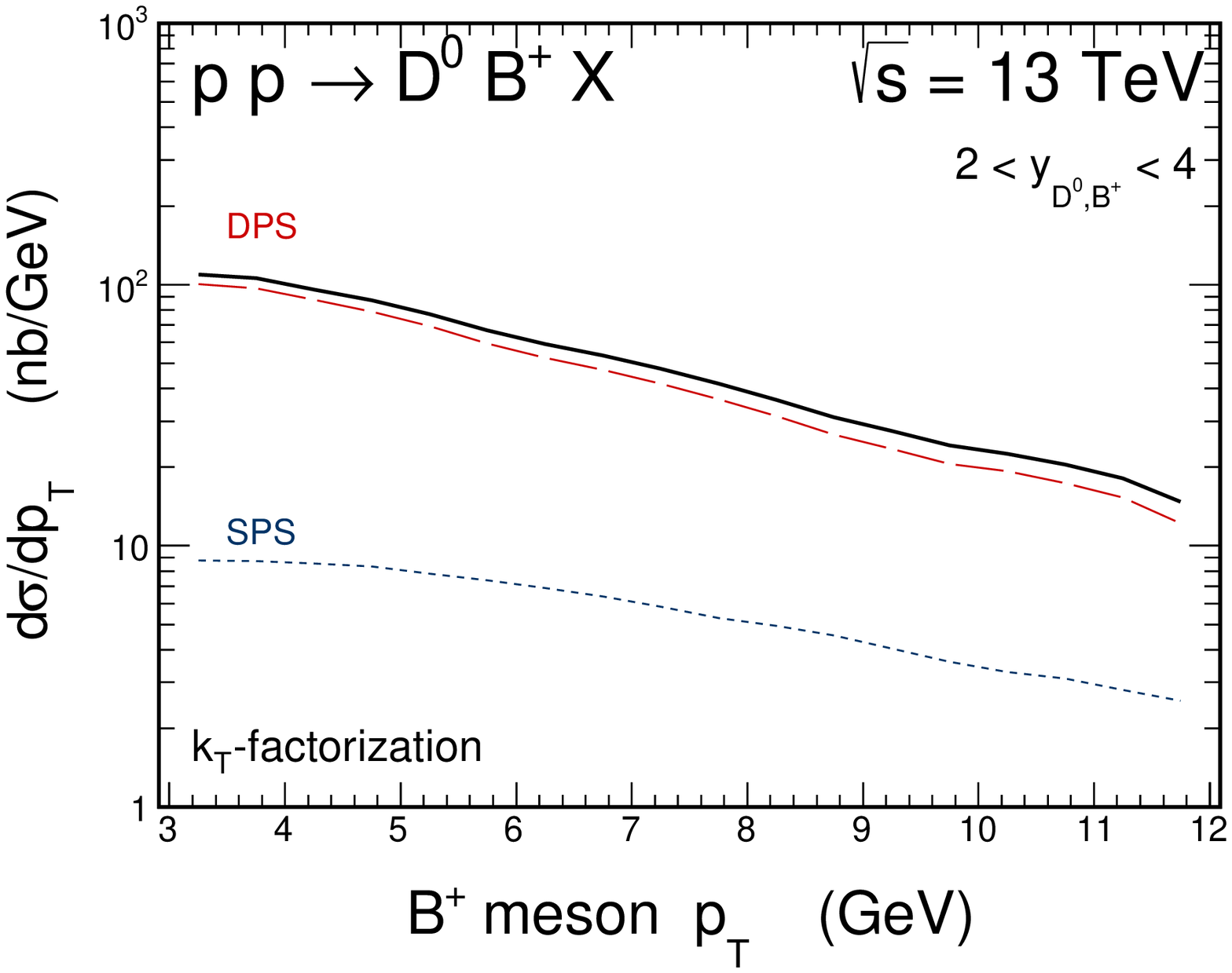}}
\end{minipage}
   \caption{
\small Transverse momentum distribution of $D^{0}$ (left) and $B^{+}$ (right) meson at $\sqrt{s} = 13$ TeV for the case of inclusive $D^{0}B^{+}$-pair production in the LHCb fiducial volume. The SPS (dotted) and the DPS (dashed) components are shown separately. The solid lines correspond to the sum of the two mechanisms under consideration. The results are obtained within the $k_{T}$-factorization approach with the KMR uPDFs. 
 }
 \label{fig:D0Bp-pT}
\end{figure}

Now we consider inclusive production of $D^{0}B^{+}$-pairs. This mode has the most advantageous $cb \to DB$ fragmentation probability and leads to the biggest cross sections. In Fig.~\ref{fig:D0Bp-pT} we show the transverse momentum distribution of $D^{0}$ (left panel) and $B^{+}$ (right panel) meson at $\sqrt{s} = 13$ TeV for the case of simultaneous $D^{0}B^{+}$-pair production in the LHCb fiducial volume defined as $2 < y < 4$ and $3 < p_{T} < 12$ GeV for both mesons. The SPS (dotted lines) and the DPS (dashed lines) components are shown separately, together with their sum (solid lines). The DPS component leads to an evident enhancement of the cross section, at the level of order of magnitude, in the whole considered kinematical domain. We predict that the $D^{0}B^{+}$ data sample, that could be collected with the LHCb detector, should be DPS dominated in the pretty much the same way as in the case of double charm production (see \textit{e.g.} Ref.~\cite{Maciula:2013kd}).    
 
\begin{figure}[!h]
\begin{minipage}{0.47\textwidth}
 \centerline{\includegraphics[width=1.0\textwidth]{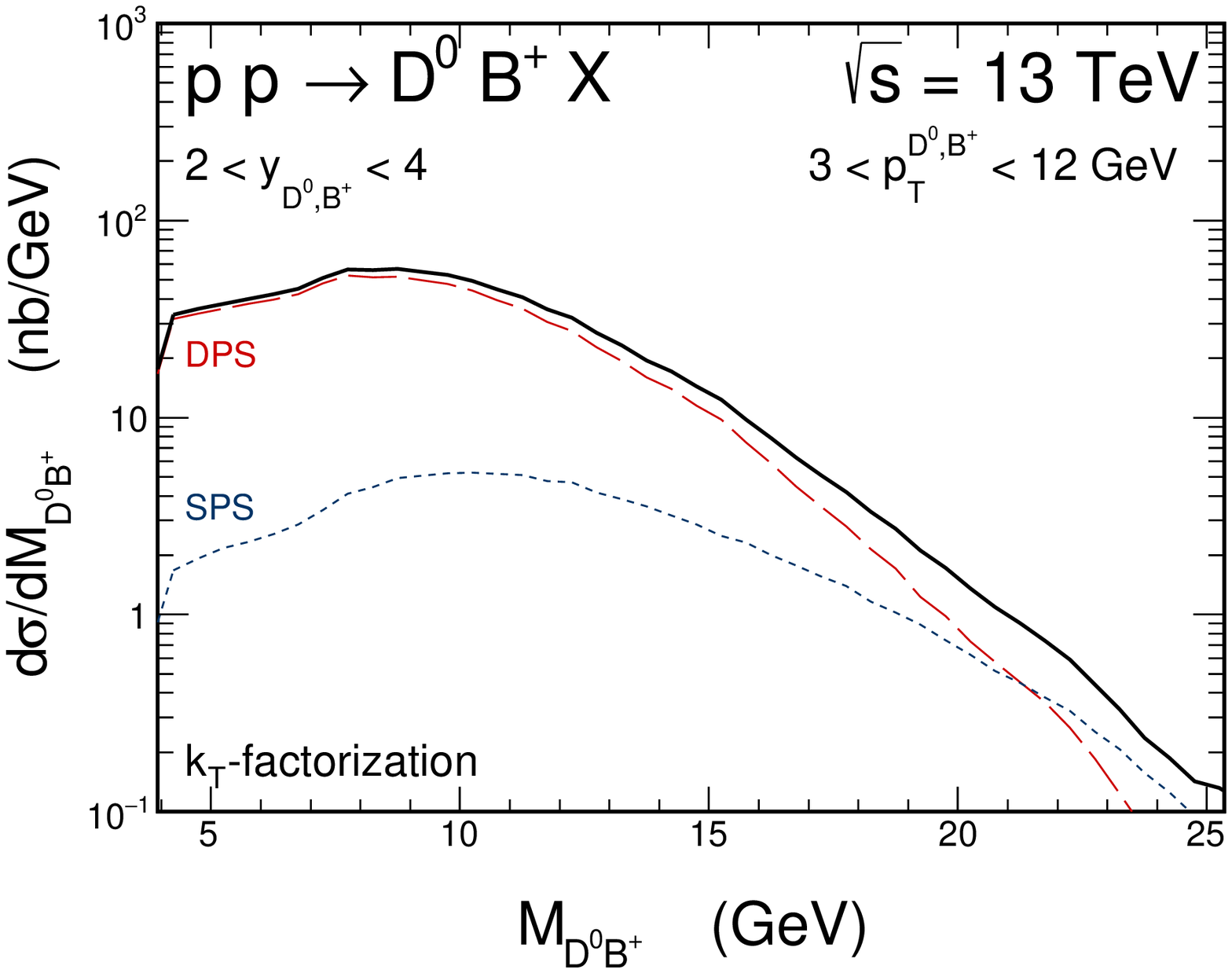}}
\end{minipage}
\hspace{0.5cm}
\begin{minipage}{0.47\textwidth}
 \centerline{\includegraphics[width=1.0\textwidth]{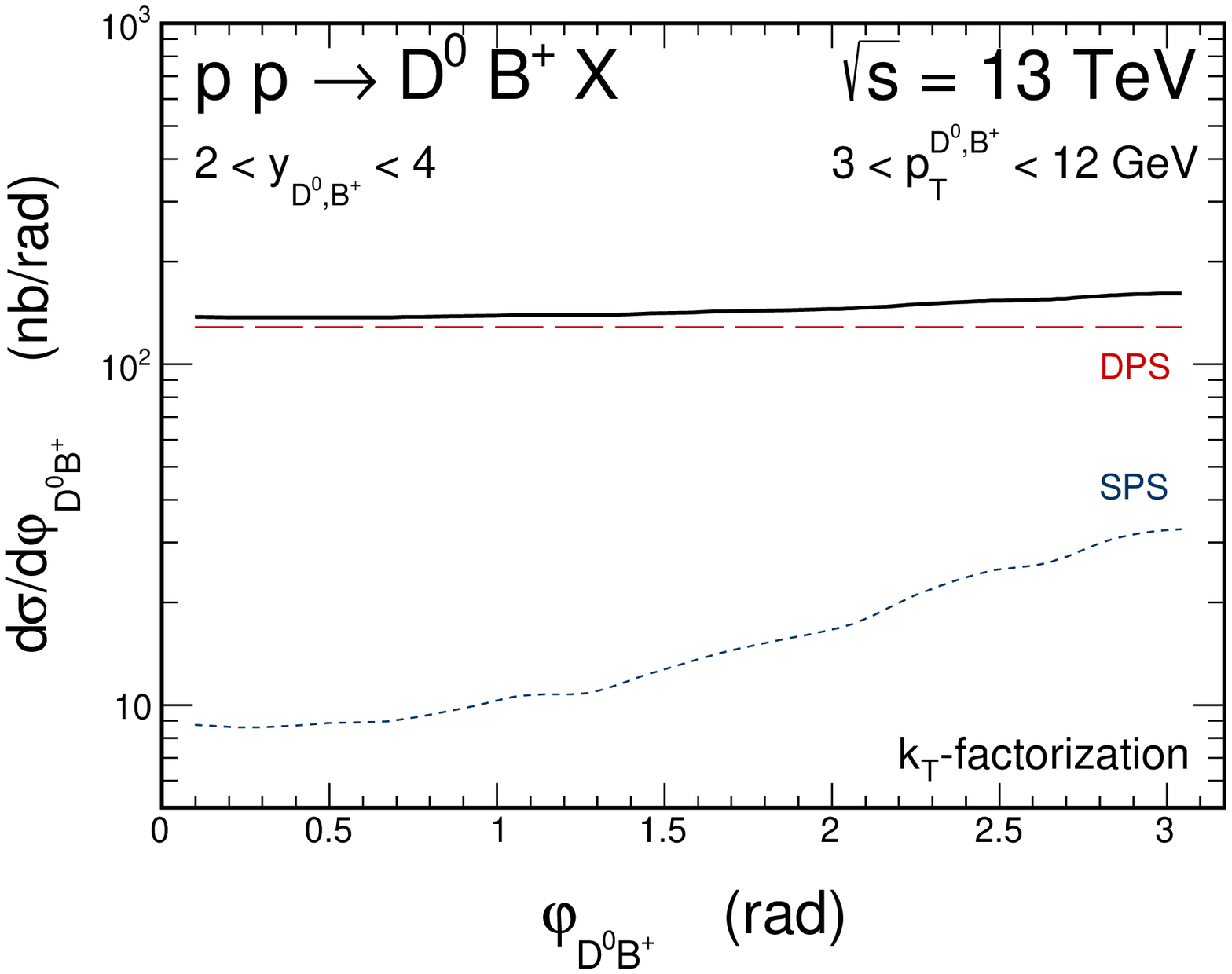}}
\end{minipage}
   \caption{
\small The same as in Fig.~\ref{fig:D0Bp-pT} but for the $D^{0}B^{+}$-pair invariant mass (left) and azimuthal angle $\varphi_{D^{0}B^{+}}$ (right) distributions.
 }
 \label{fig:D0Bp-corr}
\end{figure}

In Fig.~\ref{fig:D0Bp-corr} we present correlations observables that
could be helpful in experimental identification of the predicted DPS effects.
The characteristics of the di-meson invariant mass $M_{D^{0}B^{+}}$ 
(left panel) as well as of the azimuthal angle $\varphi_{D^{0}B^{+}}$ 
(right panel) differential distributions
is clearly determined by the large contribution of the DPS mechanism. 

\begin{figure}[!h]
\centering
\begin{minipage}{0.47\textwidth}
 \centerline{\includegraphics[width=1.0\textwidth]{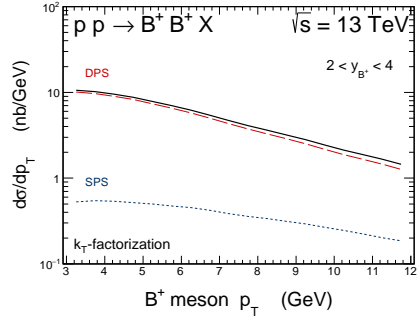}}
\end{minipage}
   \caption{
\small Transverse momentum distribution of $B^{+}$ meson at $\sqrt{s} = 13$ TeV for the case of inclusive $B^{+}B^{+}$-pair production in the LHCb fiducial volume. The SPS (dotted) and the DPS (dashed) components are shown separately. The solid lines correspond to the sum of the two mechanisms under consideration. The results are obtained within the $k_{T}$-factorization approach with the KMR uPDFs. 
 }
 \label{fig:BpBp-pT}
\end{figure}

\begin{figure}[!h]
\begin{minipage}{0.47\textwidth}
 \centerline{\includegraphics[width=1.0\textwidth]{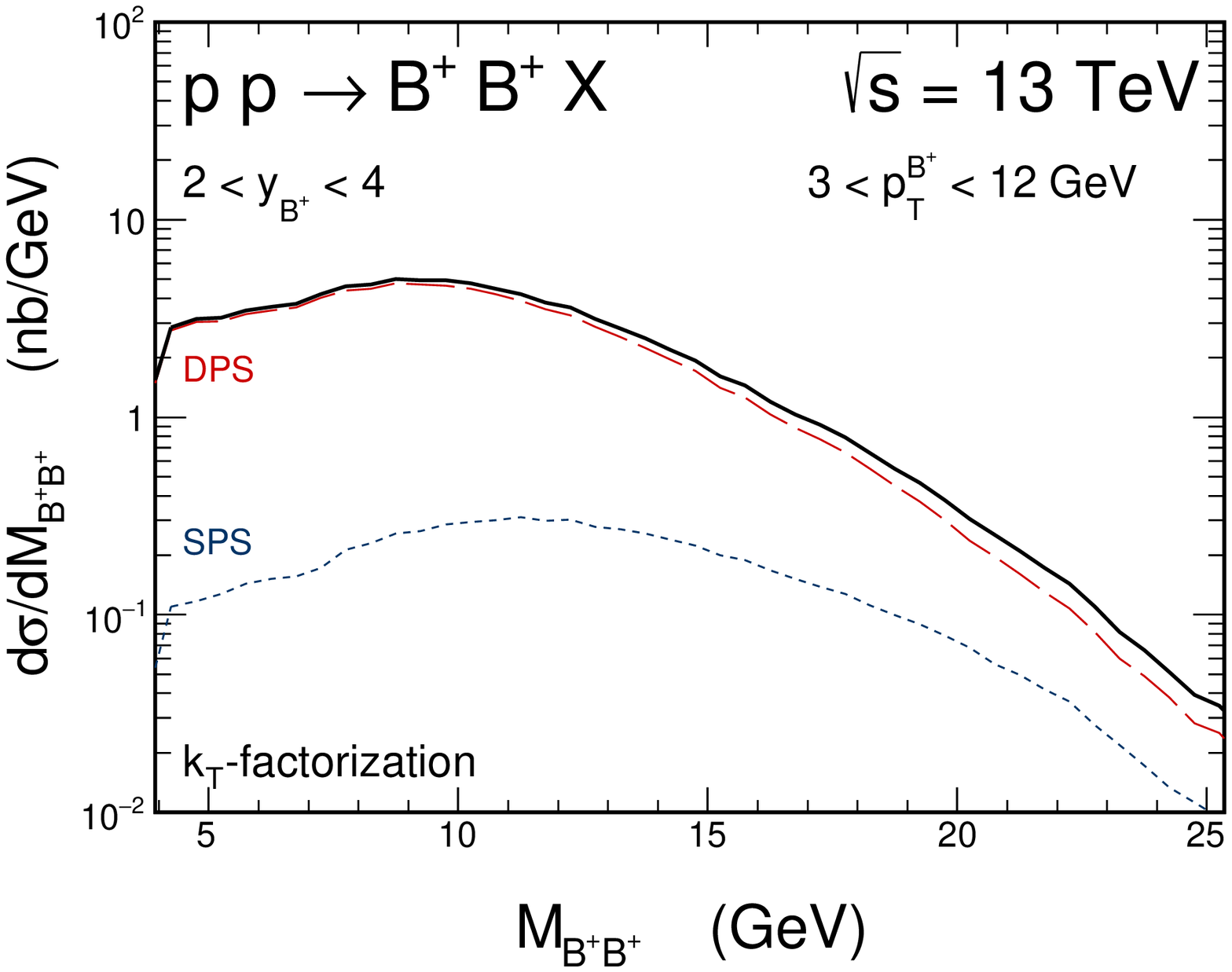}}
\end{minipage}
\hspace{0.5cm}
\begin{minipage}{0.47\textwidth}
 \centerline{\includegraphics[width=1.0\textwidth]{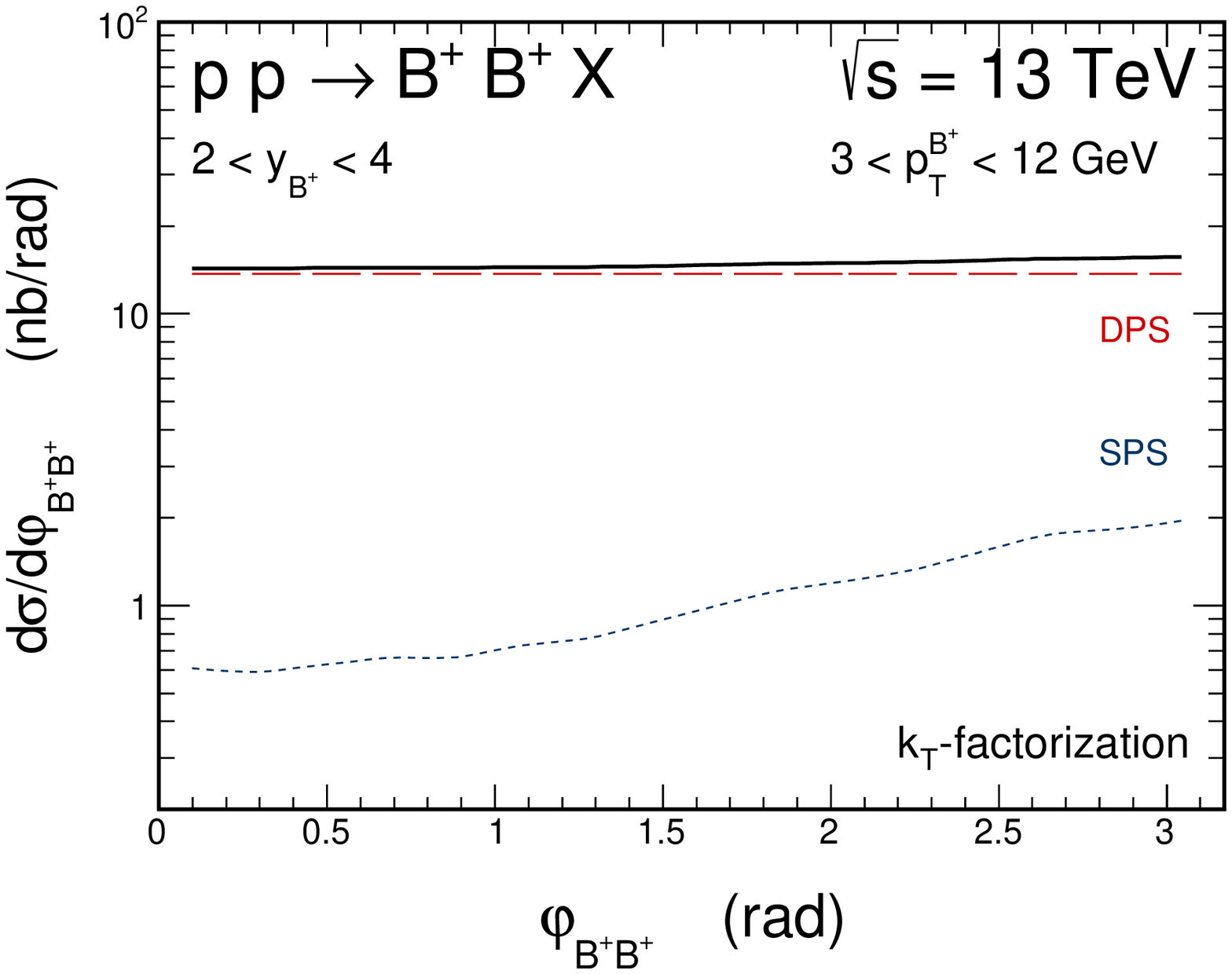}}
\end{minipage}
   \caption{
\small The same as in Fig.~\ref{fig:BpBp-pT} but for the $B^{+}B^{+}$-pair invariant mass (left) and azimuthal angle $\varphi_{B^{+}B^{+}}$ (right) distributions.
 }
 \label{fig:BpBp-Minv-Phid}
\end{figure}

The predictions for $B^+B^+$ meson-meson pair production for the LHCb
experiment leads to similar conlusions as those presented above. The effects related to 
the DPS mechanism on the $B^{+}$-meson transverse momentum 
(see Fig.~\ref{fig:BpBp-pT}),  on di-meson invariant mass 
$M_{B^{+}B^{+}}$  and on relative azimuthal angle $\varphi_{B^{+}B^{+}}$
(see left and right panels of Fig.~\ref{fig:BpBp-Minv-Phid})
distributions are pretty much the same as in the case of simultaneous
production of charm and bottom. The predicted absolute normalization of the cross section
and relative contribution of the DPS are only a bit smaller than in the $D^0B^+$ case.

To summarize the situation for the LHCb experiment, in
Table~\ref{tab:total-cross-sections}, we collect the integrated cross
sections for $D^{0}B^{+}$ and $B^{+}B^{+}$ meson-meson pair production
in nanobarns within the relevant acceptance: $2 < y_{D^{0},B^{+}} < 4$
and $3 < p_{T}^{D^{0}, B^{+}} < 12$ GeV. We predict quite large cross
sections, in particular, at $\sqrt{s}=7$ TeV the calculated cross
section for $D^{0}B^{+}$ pair production is only 5 times smaller than
the cross section already measured by the LHCb for $D^{0}D^{0}$ final
state \cite{Aaij:2012dz}. The cross sections for $B^{+}B^{+}$ are order
of magnitude smaller than in the mixed charm-bottom mode, however, still
seems measurable.  
In both cases, the DPS component is the dominant one. The relative DPS
contribution for both energies and for both experimental modes is at the
very high level of 90\%.
This makes the possible measurements a very interesting for the multi-parton interaction community.  

\begin{table}[tb]%
\caption{The integrated cross sections for $D^{0}B^{+}$ and $B^{+}B^{+}$ meson-meson pair production (in nb) within the LHCb acceptance: $2 < y_{D^{0},B^{+}} < 4$ and $3 < p_{T}^{D^{0}, B^{+}} < 12$ GeV, calculated in the $k_{T}$-factorization approach. The numbers include the charge conjugate states.}
\label{tab:total-cross-sections}
\centering %
\newcolumntype{Z}{>{\centering\arraybackslash}X}
\begin{tabularx}{1.0\linewidth}{Z Z Z Z}
\toprule[0.1em] %

Final state & Mechanism  & $\sqrt{s} = 7$ TeV  & $\sqrt{s} = 13$ TeV       \\ [-0.2ex]

\bottomrule[0.1em]

\multirow{2}{3cm}{$D^{0}B^{+} + \bar{D^{0}}B^{-}$} & DPS  & 115.50  & 418.79 \\  [-0.2ex]
 & SPS  & 21.13  & 51.46 \\  [-0.2ex]
\hline
\multirow{2}{3cm}{$B^{+}B^{+} + B^{-}B^{-}$} & DPS  & 11.04  & 43.40 \\  [-0.2ex]
 & SPS  & 1.31  & 3.39 \\  [-0.2ex]
\bottomrule[0.1em]

\end{tabularx}
\end{table}

\section{Conclusions}

In the present paper we have carefully examined
simultaneous production of charm-bottom
and bottom-bottom meson-meson pairs. It was our aim to understand the interplay of
single- and double-parton scattering processes.
The SPS results were obtained with the help of the KaTie code that allows for calculations based on $k_{T}$-factoriszation approach.
The DPS calculations have been done within the standard so far factorized
ansatz with two independent partonic scatterings.
The so-called $\sigma_{\mathrm{eff}}$ parameter have been fixed at 
the same values as used in our previous studies for double charm production.

We have considered several differential distributions for simultaneous production of
charmed and bottom mesons. The DPS mechanism have been shown to dominate
for small invariant masses of the $D B$ systems. We have predicted only
a small correlation in relative azimuthal angle, typical for DPS
dominance.

The situation for two $B^+ B^+$ meson production
is rather similar as for the mixed heavy flavour production, but here 
the dominance of the DPS over SPS is limited to smaller corners of 
the phase space.
A good description of future data will therefore require to include 
both DPS and SPS mechanisms simultaneously.
All the considered reactions should be easily measured as the
corresponding cross sections are rahter large.



\begin{thebibliography}{100}

\bibitem{Astalos:2015ivw} 
  R.~Astalos {\it et al.},
  "Proceedings of the Sixth International Workshop on Multiple Partonic Interactions at the Large Hadron Collider",
  arXiv:1506.05829 [hep-ph].

\bibitem{Proceedings:2016tff} 
  H.~Jung, D.~Treleani, M.~Strikman and N.~van Buuren,
  "Proceedings, 7th International Workshop on Multiple Partonic Interactions at the LHC (MPI@LHC 2015) : Miramare, Trieste, Italy, November 23-27, 2015",
  DESY-PROC-2016-01.

\bibitem{Luszczak:2011zp} 
  M.~{\L}uszczak, R.~Maciu{\l}a and A.~Szczurek,
  Phys.\ Rev.\ D {\bf 85}, 094034 (2012).


\bibitem{Maciula:2013kd} 
  R.~Maciu{\l}a and A.~Szczurek,
  Phys.\ Rev.\ D {\bf 87}, no. 7, 074039 (2013).

\bibitem{vanHameren:2014ava} 
  A.~van Hameren, R.~Maciu{\l}a and A.~Szczurek,
  Phys.\ Rev.\ D {\bf 89}, no. 9, 094019 (2014).

\bibitem{vanHameren:2015wva} 
  A.~van Hameren, R.~Maciu{\l}a and A.~Szczurek,
  Phys.\ Lett.\ B {\bf 748}, 167 (2015).

\bibitem{Maciula:2015vza} 
  R.~Maciu{\l}a and A.~Szczurek,
  Phys.\ Lett.\ B {\bf 749}, 57 (2015).

\bibitem{Kutak:2016ukc} 
  K.~Kutak, R.~Maciu{\l}a, M.~Serino, A.~Szczurek and A.~van Hameren,
  Phys.\ Rev.\ D {\bf 94}, no. 1, 014019 (2016).

\bibitem{Maciula:2017egq} 
  R.~Maciu{\l}a and A.~Szczurek,
  arXiv:1707.08366 [hep-ph].


\bibitem{DelFabbro:2002pw} 
  A.~Del Fabbro and D.~Treleani,
  Phys.\ Rev.\ D {\bf 66}, 074012 (2002).

\bibitem{Cazaroto:2013fua} 
  E.~R.~Cazaroto, V.~P.~Goncalves and F.~S.~Navarra,
  Phys.\ Rev.\ D {\bf 88}, no. 3, 034005 (2013).

\bibitem{Catani:1990xk} 
  S.~Catani, M.~Ciafaloni and F.~Hautmann,
  Phys.\ Lett.\ B {\bf 242}, 97 (1990).

\bibitem{Catani:1990eg} 
  S.~Catani, M.~Ciafaloni and F.~Hautmann,
  Nucl.\ Phys.\ B {\bf 366}, 135 (1991).
  
\bibitem{Bury:2015dla} 
  M.~Bury and A.~van Hameren,
  Comput.\ Phys.\ Commun.\  {\bf 196}, 592 (2015)
  [arXiv:1503.08612 [hep-ph]].

\bibitem{vanHameren:2016kkz} 
  A.~van Hameren,
  arXiv:1611.00680 [hep-ph].

\bibitem{Kimber:2001sc} 
  M.~A.~Kimber, A.~D.~Martin and M.~G.~Ryskin,
  Phys.\ Rev.\ D {\bf 63}, 114027 (2001)
  [hep-ph/0101348].
  
\bibitem{Watt:2003vf}
  G.~Watt, A.~D.~Martin and M.~G.~Ryskin,
  Phys.\ Rev.\ D {\bf 70} (2004) 014012
   Erratum: [Phys.\ Rev.\ D {\bf 70} (2004) 079902]
  [hep-ph/0309096].

\bibitem{Harland-Lang:2014zoa} 
  L.~A.~Harland-Lang, A.~D.~Martin, P.~Motylinski and R.~S.~Thorne,
  Eur.\ Phys.\ J.\ C {\bf 75}, no. 5, 204 (2015)
  [arXiv:1412.3989 [hep-ph]].

\bibitem{Peterson:1982ak} 
  C.~Peterson, D.~Schlatter, I.~Schmitt and P.~M.~Zerwas,
  Phys.\ Rev.\ D {\bf 27}, 105 (1983).

\bibitem{Diehl:2011tt}
  M.~Diehl and A.~Schafer,
  Phys.\ Lett.\ B {\bf 698} (2011) 389
  [arXiv:1102.3081 [hep-ph]].
  
\bibitem{Diehl:2011yj} 
  M.~Diehl, D.~Ostermeier and A.~Schafer,
  JHEP {\bf 1203}, 089 (2012)
  Erratum: [JHEP {\bf 1603}, 001 (2016)]
  [arXiv:1111.0910 [hep-ph]].

\bibitem{Gaunt:2009re} 
  J.~R.~Gaunt and W.~J.~Stirling,
  JHEP {\bf 1003}, 005 (2010)
  [arXiv:0910.4347 [hep-ph]].

\bibitem{Aaij:2013noa} 
  R.~Aaij {\it et al.} [LHCb Collaboration],
  J. High Energy Phys. {\bf 08}, 117 (2013).
  
\bibitem{Aaij:2012dz} 
  R.~Aaij {\it et al.} [LHCb Collaboration],
  JHEP {\bf 1206}, 141 (2012)
  Addendum: [JHEP {\bf 1403}, 108 (2014)].
  
\end{thebibliography}
\end{document}